\documentclass[12pt]{article}
\usepackage{oldgerm}
\usepackage{yfonts}
\usepackage{euler}
\usepackage{graphics}
\usepackage{bm}
\usepackage{graphicx}
\usepackage{amsmath}
\usepackage{amssymb}
\usepackage{amstext}
\usepackage{amscd}
\usepackage{amsfonts}
\usepackage{appendix}
\usepackage{wasysym}
\usepackage{epstopdf}
\usepackage{color}
\usepackage{hyperref}
\DeclareMathAlphabet{\EuFrak}{U}{euf}{m}{n}
\DeclareMathAlphabet{\EuScript}{U}{eus}{m}{n}

\newcommand{\nd}{\noindent}

\newcommand{\be}{\begin{equation}}
\newcommand{\ee}{\end{equation}}
\newcommand{\ben}{\begin{eqnarray}}
\newcommand{\een}{\end{eqnarray}}

\title{{\bf Coherent and cat states of open and closed strings}}
\author{{Mir Hameeda$^{1,2,3,a}$, Mario C. Rocca$^{4,5,6,*,b}$ }  \\
\small{$^1$ Department of Physics, Government Degree College, Tangmarg, Kashmir, 193402 India}\\
\small{$^2$ School of Physics, Damghan University, P. O. Box 3671641167, Damghan, Iran}\\
\small{$^3$ Inter University Centre for Astronomy and Astrophysics , Pune India}\\
\small{$^4$ Departamento de F\'{\i}sica,
Universidad Nacional de La Plata. Argentina}\\
\small{$^5$ Departamento de Matem\'{a}tica,
Universidad Nacional de La Plata. Argentina}\\
\small{$^6$ Consejo Nacional de Investigaciones Cient\'{\i}ficas
y Tecnol\'{o}gicas}\\
\small{(IFLP-CCT-CONICET)-C. C. 727, 1900 La Plata -
Argentina}\\
\small{\texttt{\rm{$^{a}$hme123eda@gmail.com, $^{*}$mariocarlosrocca@gmail.com}}}\\
\small{\texttt{\rm{$^{b}$rocca@fisica.unlp.edu.ar}}}}  

\date{\today}

\begin{document}

\maketitle

\vspace{-5mm}

\begin{abstract}
\nd
The covariant quantization and light cone quantization formalisms are followed to construct the coherent states of both open and closed bosonic strings. We make a systematic and straightforward use of the original definition of coherent states of harmonic oscillators to establish the coherent and their corresponding cat states. We analyze the statistics of these states by explicitly calculating the Mandel parameter and obtained interesting results about the nature of distribution of the states. A tachyonic state with imaginary mass and positive norm is obtained.\\
\nd

\end{abstract}

\newpage

\tableofcontents

\newpage

\renewcommand{\theequation}{\arabic{section}.\arabic{equation}}

\section{Introduction}

\nd
The concept of coherent states was first introduced by Schrodinger in quantum harmonic oscillators and is thus oftenly associated with harmonic oscillators. It is for the same reason that it is also known as canonical coherent states or standard coherent states and has seen application in many areas of physics. The concept of coherent states has been well explained in mathematical physics and has been applied not only in  quantization, signal processing and image processing but is backbone of quantum optics. The term coherent state was constructed by Glauber in the electromagnetic field and has been defined as the eigenstates of the annihilation operator.\cite{glau}
In Quantum mechanics coherent states being the states of minimal uncertainty has played crucial role.\cite{merz} As per their role in string theory, the concept is yet to be explored completely but till now it has been widely used for the computation of scattering amplitudes \cite{gsw}. 

\nd
The attempts to construct coherent states for a closed string in the orthonormal covariant gauge to remove the ghosts by applying the mean values of Virasoro operators have been studied long before, but the difficulties to generalize the results could not be completely eliminated \cite{cal}.

\nd
In general coherent states can be defined in two equivalent ways either as eigenstates of the annihilation operator, or as the exponential of the creation operator acting on the ground state\cite{norma}.
The concept of coherent states has been recently generalized to Lie groups \cite{lie}. The entanglement of coherent states is the current area being studied by many \cite{entangled}. But the concept of coherent states needs to be studied more in string theory to explore the wide area of utility.  Very recently the coherent states of open bosonic strings has been studied using DDF and vertex operators \cite{prev,epjc}. 
Unfortunately quantization using BRST symmetry can not be used for 
exponentially increasing states. Therefore the equivalence of our theory with that of \cite{fms} cannot be probed, unless it is 
possible to extend the quantization using BRST symmetry to exponentially increasing states. That  could be a new area of work,
clearly very general, very difficult and that escapes the purposes of our paper.
Some work on construction of coherent states has been discussed to study the cosmic superstrings \cite{cosmic}. People have attempted to establish the concept of coherent states in string theory, but no considerable theory has been established yet.\\
The concept of cat states is in its early infancy, in the well known theory of strings. In quantum mechanics this concept is attached to Schrodinger cat, which is defined as a quantum state composed of two diametrically opposed conditions at the same time. The other definition relates it to the superposition of distinct states \cite{vvv}. Cat states are actually a type of non-classical states that results through the superposition of coherent states. These are also known as even and odd coherent states\cite{vvv}. Generalization of single-mode Schrodinger cat states and the superposition of Gazeau-Klauder coherent states has been analysed to know the photon statistics. \cite{gaz,sol} \\
A straightforward way to study the photon number squeezing or photon statistics of any states is to evaluate the Mandel parameter.In well behaved coherent states like the Glauber coherent states the Mandel parameter is zero,which corresponds to the fact that the photon distribution is Poissonian. Thus depending upon the value of Mandel parameter one can easily comment on the statistics  and thus the nature of the probability distribution of the states. For the Mandel parameter greater than $0$ the statistics correspond to super Poissonian and for less than $0$, the distribution is sub-Poissonian.\cite{man} There are alternative methods to study the photon squeezing through the second order correlation function.\\
The motivation for this work is to define the true coherent  Glauber states for the bosonic string. This has not been achieved so far by a covariant formulation of the theory or by formulating it in the light cone
Inquisitiveness about the claims made by many for covariant formalism not being successful in accomplishing the task of obtaining coherent states, is the real stimulus for the present analysis.
Quite motivated by the validity of the true definition of the coherent states being an eigenfunction of the harmonic oscillator annihilation operator and having faith on the the pedagogical clout of the straight forward approach of the covariant quantization,  the present work aims at constructing the coherent states for the bosonic string. Contrary to the earlier claims, 
 we constructed and developed not only the coherent states of open and closed bosonic strings but 
have been able to remove the difficulties arising due to $g_{00}=-1$, thus were successful in obtaining the general form of coherent state. In this way we were able to define the coherent state with well satisfied physical conditions.
It is worth mentioning that instead of lack of any literature of cat states in string theory, we made a bold try and 
we calculated the complete cat states of the strings and evaluated the corresponding Mandel parameters. we have obtained very interesting and new results.
 An impressive work on tachyonic field related to complex mass fields and their forbidden propagation asymptotically as free particles and compatibility of Wheeler propagator Green function in suppressing the asymptotic free modes of the tachyonic field has been done \cite{bar3,bar4} and references therein. In this study we also obtained new tachyonic states of the string with imaginary mass and positive norm. This opens a new area to be dealt with, using Wheeler propagator Green function in our future studies.
This is first of its studies, which is very simple and straightforward and based on the original definition of the coherent states of harmonic oscillator. For the benefit of readers we have written  three appendices with the main mathematical results used in this paper.

\setcounter{equation}{0}

\section{Solution to and old problem in string theory}

As is known, the stress-energy tensor of the string must satisfy
\begin{equation}
\label{eq1.1}
T_{\alpha\beta}|phys>=0 
\end{equation}
where $|phys>$ is the physical state of the string.
That translates exactly to the Virasoro's  operators as:
\begin{equation}
\label{eq1.2}
(L_0-1)|phys>=0\
\end{equation}
\begin{equation}
\label{eq1.3}
L_m|phys>=0\;\;\;\;\;m\neq 0
\end{equation}
But this last equality cannot be satisfied for all $m$ since it leads to the cancellation of all $ L_m$. Instead it asks that:
\begin{equation}
\label{eq1.4}
(L_0-1)|phys>=0
\end{equation}
\begin{equation}
\label{eq1.5}
L_m|phys>=0\;\;\;\;\;m>0
\end{equation}
This is not equivalent to (\ref{eq1.1}). To satisfy (\ref{eq1.1}) at mean values level we ask then:
\begin{equation}
\label{eq1.6}
<[phys|(L_0-1)|phys>=0
\end{equation}
\begin{equation}
\label{eq1.7}
<phys|_m|phys>=0\;\;\;\;\;m\neq 0
\end{equation}
Which is another possible choice, more appropriate for constraints.\\

\nd In a foundational paper on coherent states for the string, Callucci \cite{cal} proves that the integral that defines the norm of the coherent state is exponentially divergent due to the $ g_ {00} $ component of the metric tensor of space by Minkowski \cite{cal}.
It also shows that the coherent states corresponding to the annihilation operators $a_n^0 $ cannot be defined by the known series for the usual coherent states.That is why it concludes that the coherent states for the string cannot be defined in a covariant way. We have managed to solve this problem using new mathematical tools developed in \cite{jpco} and references therein, based on Ultradistributions theory of Sebastiao e Silva \cite{jss, hasumi}, also known as Ultrahyperfunctions. This solution is explained in detail in subsection 5.2 of this manuscript. In Appendix A-C we show how to obtain the expression for the coherent states corresponding to the operators $a_n^0$ by a series that generalizes the Glauber series. 
We have thus obtained in a satisfactory way, the theory of coherent states in covariant form, without using DDF operators,
that do not lead to the obtaining of the true Glauber coherent states.

\setcounter{equation}{0}

\section{Review of covariant quantization}

\nd Before going for the extensive calculation of the coherent states and the corresponding cat states, we need to give a basic review of the string quantization, the commutation relations and the idea of Virasoro algebra followed in the paper. In this paper we are following the covariant quantization formalism  and later the light cone treatment to study the coherent states and their respective physicality is studied using the Virasoro constraints.  
Treating all fields $X^\mu$ as operators and imposing the constraint equations on the states is what we call covariant quantization.
$X^\mu$ and their conjugate momenta $\Pi_\mu=\frac{1}{2\pi\alpha^{'}}\dot{X_\mu}$ are promoted to operator valued fields which obey the canonical equal-time commutation relations,

\[[X^\mu(\sigma,\tau),\Pi_\nu(\sigma^{'},\tau)]=i\delta(\sigma-\sigma^{'})\delta_\nu^\mu\]
\begin{equation}
\label{2.1}
[X^\mu(\sigma,\tau),X^\nu(\sigma^{'},\tau)]=[\Pi_\mu(\sigma,\tau),\Pi_\nu(\sigma^{'},\tau)]=0
\end{equation}
These commutations relations are translated into the commutation relations for the Fourier modes $x^\mu$, $p^\mu$, $\alpha_n^\mu$ and $\tilde\alpha_n^\mu$ as

\begin{equation}
\label{2.2}
[x^\mu,p_\nu]=i\delta_\nu^\mu
\end{equation}

\begin{equation}
\label{2.3}
[\alpha_n^\mu,\alpha_m^\nu]=[\tilde\alpha_n^\mu,\tilde\alpha_m^\nu]=n\eta^{\mu\nu}\delta_{{n+m},0}
\end{equation}
The commutation relations for $x_\mu$ and $p_\mu$ are same for operators governing the position and momentum of the center of mass of the string. The commutation relations of $\alpha_n^\mu$ and $\tilde\alpha_n^\mu$ are those of harmonic oscillator creation and annihilation operators which after scaling are given as

\begin{equation}
\label{2.4}
a_n=\frac{\alpha_n}{\sqrt{n}}, n\ge 0
\end{equation}

\begin{equation}
\label{2.5}
a_n^{+}=\frac{\alpha_{-n}}{\sqrt{n}}, n\ge 0
\end{equation}
and the resulting commutation relation becomes

\begin{equation}
\label{2.6}
[a_n,a_m^{+}]=\delta_{mn}
\end{equation}
We consider a string equivalent to infinite countable number of harmonic oscillators and the matrix raising and lowering operators can be expressed as

\begin{equation}
\label{2.7}
a_n^\mu=\frac{1}{\sqrt{2}}\left(x_n^\mu+ip_n^\mu\right)
\end{equation}

\begin{equation}
\label{2.8}
a_n^{+\mu}=\frac{1}{\sqrt{2}}\left(x_n^\mu-ip_n^\mu\right)
\end{equation}
\cite{IM, BZ}.
In momentum space we define quantum number $p^\mu$ as the eigenvalue of momentum operator $\hat{p}^\mu$ we can also write the vacuum state as $|0;p>$ which obeys $\hat{p}^\mu|0,p>={p}^\mu|0,p>$\\
The elegance and convenience of the Polyakov string action makes it the most preferable choice of action, thus the starting point in any analysis of the covariant string quantization.
In its very suggestive form we write the Polyakov action as

\begin{equation}
\label{2.9}
S=-\frac{1}{4\pi\alpha^{'}}\int d\tau d\sigma\eta^{\alpha\beta}\partial_\alpha X^\mu\partial_\beta X^\nu \eta_{\mu\nu}
\end{equation}
with $\eta_{\mu\nu}$ the Minkowski metric, $\alpha, \beta$ the indices corresponding to the two world sheet coordinates $\tau$ and $\sigma$.
\cite{gsw}
The beauty of Lorentz covariant quantization of string theory lies in the fact that it treats all the string coordinates on equal footing, with all the coordinates including the $X^0$, the time coordinate become the operators, thus showing the built in time dependence of the states and in addition keeps us away from the usual Hilbert space postulates.  It has an upper hand over the light cone quantization in preserving the Lorentz symmetry , which occupies a pivotal importance in the string theory..
Keeping in view the restrictions in other quantization schemes, and the straightforward character of covariant approach, we prefer to follow the old covariant quantization formalism. In this paper we are following the covariant quantization formalism and later the light cone treatment to study the coherent states and the physicality is verified using the Virasoro constraints. All the fields and their conjugate momenta are treated as operators and imposing the constraint equations on the states is the method followed in covariant quantization.
The mode expansion for  open and closed string coordinates read as
\begin{equation}
\label{2.10}
X^{\mu}(\tau,\sigma)=x_0^\mu+\sqrt{2\alpha^{'}}\alpha_0^\mu\tau+i\sqrt{\frac{\alpha^{'}}{2}}\sum\limits_{n\neq 0}\frac{e^{-in\tau}}{n}(\alpha_n^\mu\cos{n\sigma})
\end{equation}

\begin{equation}
\label{2.11}
X^{\mu}(\tau,\sigma)=x_0^\mu+\sqrt{2\alpha^{'}}\alpha_0^\mu\tau+i\sqrt{\frac{\alpha^{'}}{2}}\sum\limits_{n\neq 0}\frac{e^{-in\tau}}{n}(\alpha_n^\mu e^{in\sigma}+\tilde\alpha_n^\mu e^{-in\sigma})
\end{equation}
\cite{gsw}
These expansions are used for all the string coordinates. For open strings only one set of modes $\alpha_n^\mu$ appears while as there are two independent sets of modes for closed string $\alpha_n^\mu$ and $\tilde\alpha_n^\mu$. The only constraints that relates the left moving and right moving modes is the level matching condition of the bosonic string. In open string, the boundary conditions force the left and right moving modes to combine into standing waves.
The Virasoro operators obtained after exploring the constraints are

\begin{equation}
\label{2.12}
L_n=\frac{1}{2}\sum\limits_p\alpha_p^\mu\alpha_{n-p,\mu}
\end{equation}

\begin{equation}
\label{2.13}
\tilde L_n=\frac{1}{2}\sum\limits_p\alpha_p^\mu\alpha_{n-p,\mu}
\end{equation}
In covariant formalism the quantum constraint satisfied for any physical state of the open string $\phi$ is
\[(L_n-\delta_{n,0})|\phi>=0 ;n\geq 0\] 
For the closed string the constraints apply to both left and right moving modes and are

\begin{equation}
\label{2.14}
(L_n-\delta_{n,0})|\phi>=0 ;n\geq 0
\end{equation}

\begin{equation}
\label{2.15}
(\tilde L_n-\delta_{n,0})|\phi>=0 ;n\geq 0
\end{equation}
The speciality of Virasoro operators is that it annihilates the physical states and thus leaves no physical state.
The commutation relations for the modes $\alpha_n^\mu$ has shown that the Virasoro generators satisfy the relation
\begin{equation}
\label{2.16}
[L_m,L_n]=(m-n)L_{m+n}+\frac{c}{12}m(m^2-1)\delta_{{m+n},0}
\end{equation}
where $c=D=26$ is the space time dimension and is proportional to the quantum effect and also called as central charge.\cite{gsw}

\section{A Useful Equality}

\setcounter{equation}{0}

A coherent state is defined as an eigenfunction of the Harmonic Oscillator annihilation operator $a$ (Glauber).\cite{glau}
Thus we have:
\begin{equation}
\label{eq3.1}
\hat{a}|\alpha>=\alpha|\alpha>
\end{equation}
{\bf (Note that for true Glauber coherent states the  eigenvalue of operator $\hat{a}$
is given by $\lambda=\lambda_\alpha=\alpha$)}
Its usual and well-known expression is
\begin{equation}
\label{eq3.2}
|\alpha>=e^{-\frac {|\alpha|^2} {2}}
\sum\limits_{m=0}^\infty\frac {(\alpha)^m} {m!}(\hat{a}^+)^m|0>
\end{equation}
Has been proved in \cite{pr1,pr2} (see also Appendix A )  that the following equality is valid:
\begin{equation}
\label{eq3.3}
|\alpha>=\left(\frac {mw} {\pi}\right)^{\frac {1} {4}}
e^{-\frac {(\alpha)^2} {2}} e^{-\frac {|\alpha|^2} {2}}
\int e^{-\frac {mw(y)^2} {2}} e^{  \sqrt{2mw}\alpha y}|y>dy
\end{equation}

As a consequence a simple component of the coherent state for the open bosonic string
is given by:
\begin{equation}
\label{eq3.4}
|\alpha_n^j>=e^{-\frac {|\alpha_n^j|^2} {2}}
\sum\limits_{m=0}^\infty\frac {(\alpha_n^j)^m} {m!}(\hat{a}_n^{+j})^m|0>
\end{equation}
Where:
\begin{equation}
\label{eq3.5}
\hat{a}_n^{+j}=\frac {\hat{\alpha}_n^j} {\sqrt{n}}\;\;\;\;;\;\;\;\;n>0\;\;\;\;;\;\;\;\;1\leq j\leq 25
\end{equation}
Taking into account (\ref{eq9.3}) we can write:
\begin{equation} 
\label{eq3.6}
|\alpha_n^j>=\left(\frac {m_n^jw_n^j} {\pi}\right)^{\frac {1} {4}}
e^{-\frac {(\alpha_n^j)^2} {2}} e^{-\frac {|\alpha_n^j|^2} {2}}
\int e^{-\frac {m_n^jw_n^j(y_n^j)^2} {2}} e^{  \sqrt{2m_n^jw_n^j}\alpha_n^j y_n^j}|y_n^j>dy_n^j
\end{equation}
This proves that either of these two expressions can be used to define the coherent state of the string for $\mu=j=1,2, ......, 25 $. The problem appears when you want to use a similar equality for the case $\mu=0$. In this case, the expression of the series that defines the coherent state component cannot be used since the integrand of the integral that defines that component is exponentially increasing. We now proceed to study and define a coherent state for the bosonic string

\setcounter{equation}{0}

\section{Coherent states for open strings}

\nd It is very well known that the annihilation operator for the
one-dimensional harmonic oscillator is given by 
\begin{equation}
\label{eq4.1}
\hat{a}_n^j=\frac {\hat{y}_n^j+i\hat{p}^j} {\sqrt{2}}
\end{equation}
In the y-representation of Quantum Mechanics, this operator is
expressed via
\begin{equation}
\label{eq4.2}
\hat{a_n}^j(y_n^j)=\frac {1} {\sqrt{2}}
\left(y_n^j+\frac {\partial} {\partial y_n^j}\right)
\end{equation}
Thus,  a coherent state is defined as the eigenfunction of the annihilator operator and indeed it has been stressed in various studies that eigenstates of the annihilation operator are the actual coherent states possessing all the required properties of coherent states because these properties are based on the commutation relation between the creation and annihilation operators of the harmonic oscillator \cite{vvv}. Thus we express the states as:
\begin{equation}
\label{eq4.3} 
\hat{a_n}^j(y_n^j)\psi_{\alpha_n^j}(y_n^j)=\frac {1} {\sqrt{2}}
\left(y_n^j\psi_{\alpha_n^j}(y_n^j)+\frac {\partial \psi_{\alpha_n^j}(y_n^j)} {\partial y_n^j}\right)=
\alpha_n^j \psi_{\alpha_n^j}(y_n^j)
\end{equation}
or, equivalently,
\begin{equation}
\label{eq4.4}
 \frac {\partial \psi_{\alpha_n^j}(y_n^j)}
{\partial y_n^j}=\left(\sqrt{2}\alpha_n^j-y_n^j\right)\psi_{\alpha_n^j}(y_n^j).
\end{equation}
The solution of this differential equation is
\begin{equation}
\label{eq4.5} 
 \psi_{\alpha_n^j}(y_n^j)=\left(\frac {1} {\pi}\right)^{\frac {1} {4}}
e^{-\frac {(\alpha_n^j)^2} {2}} e^{-\frac {|\alpha_n^j|^2} {2}}
e^{-\frac {(y_n^j)^2} {2}} e^{  \sqrt{2}\alpha_n^j y_n^j}.
\end{equation}
$j=1,2,...,25$
Thus
\begin{equation}
\label{eq4.6}    
|\alpha_n^j>=\left(\frac {1} {\pi}\right)^{\frac {1} {4}}
e^{-\frac {(\alpha_n^j)^2} {2}} e^{-\frac {|\alpha_n^j|^2} {2}}
\int e^{-\frac {(y_n^j)^2} {2}} e^{  \sqrt{2}\alpha_n^j y_n^j}|y_n^j>dy_n^j
\end{equation}
Here we used the general definition of coherent states to obtain a coherent state of the bosonic string which is expressed as:

\[|\alpha_n^j,p^j>=|\alpha_n^j>\otimes|p^j>=\left[\left(\frac {1} {\pi}\right)^{\frac {1} {4}}
e^{-\frac {(\alpha_n^j)^2} {2}} e^{-\frac {|\alpha_n^j|^2} {2}}\right.\times\]
\begin{equation}
\label{eq4.7}    
\left.\int e^{-\frac {(y_n^j)^2} {2}} e^{  \sqrt{2}\alpha_n^j y_n^j}|y_n^j>dy_n^j\right]
\otimes|p^j>
\end{equation}
And:
\[\psi_{\alpha_n^j}(y_n^j,p^j)= \psi_{\alpha_n^j}(y_n^j)\phi(p^j)=\]
\begin{equation}
\label{eq4.8} 
\left[\left(\frac {1} {\pi}\right)^{\frac {1} {4}}
e^{-\frac {(\alpha_n^j)^2} {2}} e^{-\frac {|\alpha_n^j|^2} {2}}
e^{-\frac {(y_n^j)^2} {2}} e^{  \sqrt{2}\alpha_n^j y_n^j}\right]\phi(p^j).
\end{equation}
Now in order for the state obtained above to be an admissible or a representative of a physical state we need to prove the constraint relations imposed by Virasoro operators on the strings. Here we analyze the physicality of states of open strings. 
In this case we will demand that they are verified in average value (See (\ref{eq4.22}),(\ref{eq4.23})). We have for Virasoro operator:
\begin{equation}
\label{eq4.9}
L_0=\frac {1} {2}\hat{\alpha}_0^2+\sum\limits_{s=1}^{\infty}
\hat{\alpha}_{-s}\cdot\hat{\alpha}_s
\end{equation}

\begin{equation}
\label{eq4.10}
L_k=\frac {1} {2}\sum\limits_{s=-\infty}^{\infty}
\hat{\alpha}_{k-s}\cdot\hat{\alpha}_s
\end{equation}
And for annihilation and creation operators:
\begin{equation}
\label{eq4.11}
\hat{a}_n^j=\frac {1} {\sqrt{n}}\hat{\alpha}_n^j\;\;\;n>0
\end{equation}

\begin{equation}
\label{eq4.12}
\hat{a}^{+j}_n=\frac {1} {\sqrt{n}}\hat{\alpha}_{-n}^j\;\;\;n>0
\end{equation}
Thus we get in the $|y>$ representation
\begin{equation}
\label{eq4.13}
L_0=\frac {\alpha_0^2} {2}+\sum\limits_{j=1}^3
\sum\limits_{s=1}^{\infty}s\frac {\left(y_s^j-\frac {\partial} {\partial y_{s}^j}\right)}
{\sqrt{2}}
\frac {\left(y_n^j+\frac {\partial} {\partial y_s^j}\right)}
{\sqrt{2}}+\alpha_{-s}^0\alpha_{s0}
\end{equation}

Now we consider the case $\mu=0$. For it we have: 

\begin{equation}
\label{eq4.14}
\hat{a}_n^0=\frac {\hat{y}_n^0-i\hat{p}^0} {\sqrt{2 }}
\end{equation}

The coherent state is now:
\begin{equation}
\label{eq4.15} 
 \psi_{\alpha_n^0}(y_n^0)=\left(\frac {1} {\pi}\right)^{\frac {1} {4}}
e^{\frac {(\alpha_n^0)^2} {2}} e^{\frac {|\alpha_n^0|^2} {2}}
e^{\frac {(y_n^0)^2} {2}} e^{- \sqrt{2}\alpha_n^0 y_n^0}.
\end{equation}
And:
\begin{equation}
\label{eq4.16}    
|\alpha_n^0>=\left(\frac {1} {\pi}\right)^{\frac {1} {4}}
e^{\frac {(\alpha_n^0)^2} {2}} e^{\frac {|\alpha_n^0|^2} {2}}
\int e^{\frac {(y_n^0)^2} {2}} e^{-\sqrt{2}\alpha_n^0 y_n^0}|y_n^0>dy_n^0
\end{equation}
We now calculate its norm:

\[<\alpha_n^0|\alpha_n^0>=\frac {1} {\pi^{\frac {1} {2}}}\int e^{\frac {x^2} {2}}e^{\frac {y^2} {2}}
e^{2\alpha_{Rn}^{02}}e^{-\sqrt{2}\alpha_{Rn}^0(x+y)}
e^{-\sqrt{2}\alpha_{In}^j(x-y)}<x|y>dx dy =\]
\[\frac {1} {\pi^{\frac {1} {2}}}\int e^{\frac {x^2} {2}}e^{\frac {y^2} {2}}
e^{2\alpha_{Rn}^{02}}e^{-\sqrt{2}\alpha_{Rn}^0(x+y)}
\delta(x-y)dx dy =\]
\[\frac {1} {\pi^{\frac {1} {2}}}\int e^{x^2}
e^{2\alpha_{Rn}^{02}}e^{-2\sqrt{2}\alpha_{Rn}^0x} dx =\]
\[\frac {1} {\pi^{\frac {1} {2}}}\int
e^{(x- {\sqrt{2}}{\alpha_{Rn}^0})^2} dx=\]
\[\frac {1} {\pi^{\frac {1} {2}}}\int
e^{z^2}dz=\]
\begin{equation}
\label{eq4.17}
\frac {1} {\pi^{\frac {1} {2}}}\int\limits_0^\infty s^{-3/2} e^{\frac {1} {s}}ds
\end{equation}
Now we use the result \cite{gr1}(Appendix B):
\begin{equation}
\label{eq4.18}
\int\limits_0^{\infty}x^{\nu-1}
e^{\frac {\beta} {x}}dx=\cos(\pi\nu)\beta^{\nu}
\Gamma(-\nu),
\end{equation}
As a consequence:

\begin{equation}
\label{eq4.19}    
|||\alpha_n^0>||=0
\end{equation}

Thus the complete state is defined as:

\begin{equation}
\label{eq4.20}
|\alpha,p>=\prod\limits_{n\neq 0}\prod\limits_{j=1}^{25}\otimes|\alpha_n^j>\otimes|\alpha_n^0>\otimes|p>
\end{equation}
Since the coherent state has a null norm, that is
\begin{equation}
\label{eq4.21}
<\alpha,p^{'}|\alpha,p>=0
\end{equation}
It is then verified that
\begin{equation}
\label{eq4.22}
<\alpha,p^{'}|(L_0-I)|\alpha,p>=0
\end{equation}
and
\begin{equation}
\label{eq4.23}
<\alpha,p^{'}|L_k|\alpha,p>=0\;\;\;\;;\;\;\;\;k>0
\end{equation}
We must clarify that in this case the constraints of Virasoro can only be satisfied in average value.
The mass of the coherent state is given by:
\begin{equation}
\label{eq4.24}
M^2\delta(p-p^{'})=<\alpha,p^{'}|2L_0-2-\alpha_0^2|\alpha,p>=0
\end{equation}
As a consequence:
\begin{equation}
\label{eq4.25}
M^2=0
\end{equation}
We will prove that:
\begin{equation}
\label{eq4.26}
I=\int |\alpha_n^j>\frac {d\alpha_n^j} {\pi}<\alpha_n^j|
\end{equation}
and 
\begin{equation}
\label{eq4.27}
0=\int |\alpha_n^0>\frac {d\alpha_n^0} {\pi}<\alpha_n^0|
\end{equation}
Let $\alpha_n^j$ be given by $\alpha_n^j=\alpha_{Rn}^j+i\alpha_{In}^j$
Then we have
\[\int |\alpha_n^j>\frac {d\alpha_n^j} {\pi}<\alpha_n^j|=\]
\[\frac {1} {\pi^{\frac {3} {2}}}\int
e^{-\frac {x^2} {2}}e^{-\frac {y^2} {2}}
e^{-2\alpha_{Rn}^{j2}}e^{\sqrt{2}\alpha_{Rn}^j(x+y)}
e^{\sqrt{2}\alpha_{In}^j(x-y)}|x><y|dx dy d\alpha_{Rn}^j d\alpha_{In}^j=\]
\[\frac {1} {\pi^{\frac {1} {2}}}\sqrt{2}\int
e^{-\frac {x^2} {2}}e^{-\frac {y^2} {2}}
e^{-2\alpha_{Rn}^{j2}}e^{\sqrt{2}\alpha_{Rn}^j(x+y)}
\delta(x-y)|x><y|dx dy d\alpha_{Rn}^j=\]
\[\frac {1} {\pi^{\frac {1} {2}}}\sqrt{2}\int e^{-x^2}
e^{-2\alpha_{Rn}^{j2}}e^{2\sqrt{2}\alpha_{Rn}^jx}
|x><x|dx d\alpha_{Rn}^j=\]
\[\frac {1} {\pi^{\frac {1} {2}}}\sqrt{2}\int
e^{-2(\alpha_{Rn}^j-{\frac {x} {\sqrt{2}})^2}}
|x><x|dx d\alpha_{Rn}^j=\]
\begin{equation}
\label{eq4.28}
\int |x>dx<x|=I 
\end{equation}
Let now $\alpha_n^0$ be given by $\alpha_n^0=\alpha_{Rn}^0+i\alpha_{In}^0$
\[\int |\alpha_n^0>\frac {d\alpha_n^0} {\pi}<\alpha_n^0|=\]
\[\frac {1} {\pi^{\frac {3} {2}}}\int e^{\frac {x^2} {2}}e^{\frac {y^2} {2}}
e^{2\alpha_{Rn}^{02}}e^{-\sqrt{2}\alpha_{Rn}^0(x+y)}
e^{-\sqrt{2}\alpha_{In}^j(x-y)}|x><y|dx dy d\alpha_{Rn}^0 d\alpha_{In}^0=\]
\[\frac {1} {\pi^{\frac {1} {2}}}\sqrt{2}\int e^{\frac {x^2} {2}}e^{\frac {y^2} {2}}
e^{2\alpha_{Rn}^{02}}e^{-\sqrt{2}\alpha_{Rn}^0(x+y)}
\delta(x-y)|x><y|dx dy d\alpha_{Rn}^0=\]
\[\frac {1} {\pi^{\frac {1} {2}}}\sqrt{2}\int e^{x^2}
e^{2\alpha_{Rn}^{02}}e^{-2\sqrt{2}\alpha_{Rn}^0x}
|x><x|dx d\alpha_{Rn}^0=\]
\[\frac {1} {\pi^{\frac {1} {2}}}\sqrt{2}\int
e^{2(\alpha_{Rn}^0-{\frac {x} {\sqrt{2}})^2}}
|x><x|dx d\alpha_{Rn}^0=\]
\[\frac {1} {\pi^{\frac {1} {2}}}\sqrt{2}\int
e^{2z^2}dz\int
|x>dx<x|=\]
\begin{equation}
\label{eq4.29}
0\int |x>dx<x|=0 
\end{equation}
As a consequence we conclude that the components with $\mu=0$ can be eliminated from the coherent state and the coherent states are redefined as
\begin{equation}
\label{eq4.30}
|\alpha,p>=\prod\limits_{n\neq 0}\prod\limits_{j=1}^{25}\otimes|\alpha_n^j>\otimes|p>
\end{equation}
Thus the redefined form of coherent states is with us and we then have:
\begin{equation} 
\label{eq4.31}
<\alpha,p^{'}|(L_0-I)|\alpha,p>=\left(\frac {p^2} {2}-1+\sum\limits_{n=1}^\infty n|\alpha_n|^2\right)
\delta(p-p^{'})=0
\end{equation} 
and as a consequence:
\begin{equation} 
\label{eq4.32}
\frac {p^2} {2}-1+\sum\limits_{n=1}^\infty n|\alpha_n|^2=0
\end{equation}  
When $k>0$ they must satisfy:
\[ <\alpha,p^{'}|L_k|\alpha,p>=<\alpha,p^{'}|\left(
\sum\limits_{n=0}^k\sqrt{k-n}\sqrt{n}\hat{a}_n\hat{a}_{k-n}+\right.\]
\begin{equation}
\label{eq4.33}
\left.+\sum\limits_{n=k+1}^\infty\sqrt{n-k}\sqrt{n}\hat{a}_n\hat{a}_{n-k}^+
\sum\limits_{n=1}^\infty \sqrt{n+k}\sqrt{n}\hat{a}_{k+n}\hat{a}_{n}\right)|\alpha,p>=0
\end{equation}
And when $k<0$ :
\[ <\alpha,p^{'}|L_k|\alpha,p>=<\alpha,p^{'}|\left(
\sum\limits_{n=0}^\infty\sqrt{n-k}\sqrt{n}\hat{a}_{n-k}^+\hat{a}_{n}+\right.\]
\begin{equation}
\label{eq4.34}
\left.+\sum\limits_{n=-k}^\infty\sqrt{n+k}\sqrt{n}\hat{a}_n^+\hat{a}_{n+k}+
\sum\limits_{n=1}^{-k-1}\sqrt{-n-k}\sqrt{n}\hat{a}_{-k-n}^+\hat{a}_{n}^+\right)|\alpha,p>=0
\end{equation}
The mass of the coherent state is now:
\begin{equation}
\label{eq4.35}
M^2\delta(p^{'}-p)=<\alpha,p^{'}|(2L_0-2-\alpha_0^2)|\alpha,p>
\end{equation}
And then:
\begin{equation}
\label{eq4.36}
M^2=2\sum\limits_{n=1}^{\infty}n|\alpha_n|^2-2
\end{equation}
{\bf Looking closely at (\ref{eq4.36}) we can see that by suitably choosing the $\alpha_n$ we can obtain states for which $M^2<0$ and $ <\alpha,p^{'}|\alpha,p>=\delta(p^{'}-p)$ that is, a tachyonic state whose norm is positive. For example if we select
$\alpha=0$.}

In the section 11 of the paper, we will prove that these results coincide with those obtained by defining the coherent states in the Light cone quantization formalism.

\setcounter{equation}{0}

\section{Mandel Parameter for coherent states of open bosonic strings}

The definition of Mandel parameter \cite{man} $Q$ for a single mode of the string is given as $Q=Q_1+Q_2$ where:
\begin{equation}
\label{eq5.1} 
Q_1=-\frac{<\hat{a}^{+j}_n\hat{a}_n^j>-<(\hat{a}^{+j}_n\hat{a}_n^j\hat{a}^{+j}_n\hat{a}_n^j)>+<\hat{a}^{+j}_n\hat{a}_n^j>^2}{<\hat{a}^{+j}_n\hat{a}_n^j>}
\end{equation}
(We use the convention summation  repeated indices for $1\leq j\leq 25$)
We use this expression for Mandel parameter to evaluate the Mandel parameter of the coherent states of open strings obtained in the previous section. We evaluate the terms as below

\begin{equation}
\label{eq5.2} 
<\hat{a}^{+j}_n\hat{a}_n^j>=<\alpha_n^{j*}|\hat{a}^{+j}_n\hat{a}_n^j|\alpha_n^{j}>
\end{equation}
Now $\hat{a}_n^j|\alpha_n^j>=\alpha_n^j|\alpha_n^j>$ and as consequence:
$<\alpha_n^{*j}|\hat{a}_n^{+j}=<\alpha_n^{*j}|\alpha_n^{*j}$.Thus
\begin{equation}
\label{eq5.3} 
<\hat{a}^{+j}_n\hat{a}_n^j>=|\alpha_n^j|^2
\end{equation}
Thus substituting the results in the equation for $Q$ parameter we could get the Mandel parameter as:
\begin{equation}
\label{eq5.4} 
Q_1=-\frac {|\alpha_n^j|^2-|\alpha_n^j|^2<\hat{a}_n^j\hat{a}^{+j}_n>+|\alpha_n^j|^4}{|\alpha_n^j|^2}
\end{equation}
With $<\hat{a}_n^j\hat{a}^{+j}_n>=<1+\hat{a}^{+j}_n\hat{a}_n^j>=1+|\alpha_n^j|^2$
We obtain
\begin{equation}
\label{eq5.5} 
Q_1=0
\end{equation}
Similarly,for $\mu=0$ we have:

\begin{equation}
\label{eq5.6} 
Q_2=-\frac{<\hat{a}^{+0}_n\hat{a}_n^0>-<(\hat{a}^{+0}_n\hat{a}_n^0\hat{a}^{+0}_n\hat{a}_n^0)>+<\hat{a}^{+0}_n\hat{a}_n^0>^2}
{<\hat{a}^{+0}_n\hat{a}_n^0>}
\end{equation}
Again we calculate the terms and use then for calculating $Q$ as

\begin{equation}
\label{eq5.7} 
<\hat{a}^{+0}_n\hat{a}_n^0>=<\alpha_n^{0*}|\hat{a}^{+0}_n\hat{a}_n^0|\alpha_n^{0}>
\end{equation}
Now $\hat{a}_n^0|\alpha_n^0>=\alpha_n^0|\alpha_n^0>$ and as consequence:
$<\alpha_n^{*0}|\hat{a}_n^{+0}=<\alpha_n^{*0}|\alpha_n^{*0}$.Thus
\begin{equation}
\label{eq5.8} 
<\hat{a}^{+0}_n\hat{a}_n^0>=|\alpha_n^0|^2
\end{equation}
Then we have:
\begin{equation}
\label{eq5.9} 
Q_2=-\frac {|\alpha_n^0|^2-|\alpha_n^0|^2<\hat{a}_n^0\hat{a}^{+0}_n>+|\alpha_n^0|^4}{|\alpha_n^0|^2}
\end{equation}
With $<\hat{a}_n^0\hat{a}^{+0}_n>=<-1+\hat{a}^{+0}_n\hat{a}_n^0>=-1+|\alpha_n^0|^2$
We obtain
\begin{equation}
\label{eq5.10} 
Q_2=-2
\end{equation}
Then:
\begin{equation}
\label{eq5.11} 
Q=-2
\end{equation}
$Q$ indicates the sub-Poisson nature of distribution of states.
For the redefined coherent state we have
\begin{equation}
\label{eq5.12} 
Q=0
\end{equation}
Distribution of states is Poisson.

\setcounter{equation}{0}

\section{Mandel parameter for even cat states of open strings}

The cat state is defined as the quantum superposition of two opposite phases coherent states of a single optical mode. We define even cat state as
\begin{equation}
\label{eq6.1}
|Cat_e(\alpha_n^j)>=C(|\alpha_n^j>+|-\alpha_n^j>)
\end{equation}
Its norm is given by:
\[<Cat_e(\alpha_n^{*j})|Cat_e(\alpha_n^j>=\]
\begin{equation}
\label{eq6.2}
|C|^2 [<\alpha_n^{*j}|\alpha_n^j>+<\alpha_n^{*j}|-\alpha_n^j>+<-\alpha_n^{*j}|\alpha_n^j>+<-\alpha_n^{*j}|-\alpha_n^j>]
\end{equation}
Or equivalently:
\begin{equation}
\label{eq6.3}
<Cat_e(\alpha_n^{*j})|Cat_e(\alpha_n^j>=|C|^2 [2+2\Re[<\alpha_n^{*j}|-\alpha_n^j>]]
\end{equation}
As we have:
\begin{equation}
\label{eq6.4}
<\alpha_n^{*j}|-\alpha_n^j>=e^{-2|\alpha_N^J|^2}
\end{equation}
We obtain:
\begin{equation}
\label{eq6.5}
<Cat_e(\alpha_n^{*j})|Cat_e(\alpha_n^j>=|C|^22[1+e^{-2|\alpha_N^J|^2}]
\end{equation}
And as a consequence:
\begin{equation}
\label{eq6.6}
C=\frac {e^{|\alpha_N^J|^2}} {\sqrt{2(1+e^{2|\alpha_N^J|^2})}}
\end{equation}
For this cat state it is verified:
\begin{equation}
\label{eq6.7}
a_n^j|Cat_e(\alpha_n^j)>=\alpha_n^j\frac {e^{|\alpha_N^J|^2}} {\sqrt{2(1+e^{2|\alpha_N^J|^2})}}(|\alpha_n^j>-|-\alpha_n^j>)
\end{equation}
Also we have:
\begin{equation}
\label{eq6.8}
<a_n^{+j}a_n^j>=|\alpha_n^j|^2\frac {e^{2|\alpha_N^J|^2}-1} {e^{2|\alpha_N^J|^2}+1}
\end{equation}
And:
\begin{equation}
\label{eq6.9}
<a_n^{+j}a_n^ja_n^{+j}a_n^j>=|\alpha_n^j|^2\frac {e^{2|\alpha_N^J|^2}-1} {e^{2|\alpha_N^J|^2}+1}\left(1
+|\alpha_n^j|^2\frac {e^{2|\alpha_N^J|^2}+1} {e^{2|\alpha_N^J|^2}-1}\right)
\end{equation}
Thus we obtain:
\begin{equation}
\label{eq6.10}
<a_n^{+j}a_n^ja_n^{+j}a_n^j>=|\alpha_n^j|^2\frac {e^{2|\alpha_N^J|^2}-1} {e^{2|\alpha_N^J|^2}+1}
+|\alpha_n^j|^4
\end{equation}
As a consequence:
\begin{equation}
\label{eq6.11}
Q_1=4|\alpha_n^j|^2\frac {e^{2|\alpha_N^J|^2}} {e^{4|\alpha_N^J|^2}-1}
\end{equation}
Or equivalently:
\begin{equation}
\label{eq6.12}
Q_1=\frac {2|\alpha_n^j|^2} {\sinh(2|\alpha_n^j|^2)}
\end{equation}

For $\mu=0$ we have:$Q_2=-2$
Thus
\begin{equation}
\label{eq6.13}
Q=\frac {2|\alpha_n^j|^2} {\sinh(2|\alpha_n^j|^2)}-2
\end{equation}
For the redefined cat state we have
\begin{equation}
\label{eq6.14}
Q=\frac {2|\alpha_n^j|^2} {\sinh(2|\alpha_n^j|^2)}
\end{equation}
The Mandel parameter of cat states show clear dependence on $|\alpha_n^j|$. The nature of the distribution, whether it is Poissonian, sub-Poissonian or super-Poissonian will be decided by the value of $|\alpha_n^j|$.

\setcounter{equation}{0}

\section{Mandel parameter for odd cat states of open strings}

\setcounter{equation}{0}

The whole calculation of Mandel parameter for even cat states is repeated for the odd cat states. We define odd cat state as
\begin{equation}
\label{eq7.1}
|Cat_o(\alpha_n^j)>=C(|\alpha_n^j>-|-\alpha_n^j>)
\end{equation}
The norm of this state is:
\[<Cat_o(\alpha_n^{*j})|Cat_o(\alpha_n^j>=\]
\begin{equation}
\label{eq7.2}
|C|^2 [<\alpha_n^{*j}|\alpha_n^j>-<\alpha_n^{*j}|-\alpha_n^j>+<-\alpha_n^{*j}|\alpha_n^j>+<-\alpha_n^{*j}|-\alpha_n^j>]
\end{equation}
Or equivalently:
\begin{equation}
\label{eq7.3}
<Cat_o(\alpha_n^{*j})|Cat_o(\alpha_n^j>=|C|^2 [2-2\Re[<\alpha_n^{*j}|-\alpha_n^j>]]
\end{equation}
As is verified:
\begin{equation}
\label{eq7.4}
<\alpha_n^{*j}|-\alpha_n^j>=e^{-2|\alpha_N^J|^2}
\end{equation}
Or:
\begin{equation}
\label{eq7.5}
<Cat_o(\alpha_n^{*j})|Cat_o(\alpha_n^j>=|C|^22[1-e^{-2|\alpha_N^J|^2}]
\end{equation}
As a consequence:
\begin{equation}
\label{eq7.6}
C=\frac {e^{|\alpha_N^J|^2}} {\sqrt{2(e^{2|\alpha_N^J|^2}-1)}}
\end{equation}
We have now:
\begin{equation}
\label{eq7.7}
a_n^j|Cat_o(\alpha_n^j)>=\alpha_n^j\frac {e^{|\alpha_N^J|^2}} {\sqrt{2(e^{2|\alpha_N^J|^2}-1)}}(|\alpha_n^j>+|-\alpha_n^j>)
\end{equation}
And then:
\begin{equation}
\label{eq7.8}
<a_n^{+j}a_n^j>=|\alpha_n^j|^2\frac {e^{2|\alpha_N^J|^2}+1} {e^{2|\alpha_N^J|^2}-1}
\end{equation}
Moreover:
\begin{equation}
\label{eq7.9}
<a_n^{+j}a_n^ja_n^{+j}a_n^j>=|\alpha_n^j|^2\frac {e^{2|\alpha_N^J|^2}+1} {e^{2|\alpha_N^J|^2}-1}\left(1
+|\alpha_n^j|^2\frac {e^{2|\alpha_N^J|^2}-1} {e^{2|\alpha_N^J|^2}+1}\right)
\end{equation}
Or equivalently:
\begin{equation}
\label{eq7.10}
<a_n^{+j}a_n^ja_n^{+j}a_n^j>=|\alpha_n^j|^2\frac {e^{2|\alpha_N^J|^2}+1} {e^{2|\alpha_N^J|^2}-1}
+|\alpha_n^j|^4
\end{equation}
Then, we have the result:
\begin{equation}
\label{eq7.11}
Q_1=-4|\alpha_n^j|^2\frac {e^{2|\alpha_N^J|^2}} {e^{4|\alpha_N^J|^2}-1}
\end{equation}
And finally:
\begin{equation}
\label{eq7.12}
Q_1=-\frac {2|\alpha_n^j|^2} {\sinh(2|\alpha_n^j|^2)}
\end{equation}
For $\mu=0$ we obtain:$Q_2=-2$.
Thus
\begin{equation}
\label{eq7.13}
Q=-\frac {2|\alpha_n^j|^2} {\sinh(2|\alpha_n^j|^2)}-2
\end{equation}
For the redefined cat states we have:
\begin{equation}
\label{eq7.14}
Q=-\frac {2|\alpha_n^j|^2} {\sinh(2|\alpha_n^j|^2)}
\end{equation}

\nd The results Mandal parameter for odd cat states are interpreted the same way as that of the even states.

\section{Coherent states for closed strings}

\setcounter{equation}{0}
The mathematical analysis of closed strings involves a doubling of the degrees of freedom of that of the open strings. Here the left or right moving modes are the same as the standing waves on open strings. In string theory, closed strings are of particular importance because the spectrum includes a massless graviton. The two sets of modes are independent except for the relation $L_0=\tilde L_0$ \cite{gsw}
The covariant quantization of the closed strings in fact brings no complications. The procedure and the general behavior is very similar to that of open strings. Here we have two sets of covariant Virasoro operators and the operators which possess non-negative mode number annihilate the physical states. In fact the left and right number operators have the same eigenvalue when acting on physical states. We begin by considering the coherent states:
\begin{equation}
\label{eq8.1} 
 \psi_{\alpha_n^j}(y_n^j)=\left(\frac {1} {\pi}\right)^{\frac {1} {4}}
e^{-\frac {(\alpha_n^j)^2} {2}} e^{-\frac {|\alpha_n^j|^2} {2}}
e^{-\frac {(y_n^j)^2} {2}} e^{  \sqrt{2}\alpha_n^j y_n^j}.
\end{equation}
\begin{equation}
\label{eq8.2}    
\tilde\psi_{\tilde\alpha_n^j}(\tilde{y}_n^j)=\left(\frac {1} {\pi}\right)^{\frac {1} {4}}
e^{-\frac {(\tilde\alpha_n^j)^2} {2}} e^{-\frac {|\tilde\alpha_n^j|^2} {2}}
e^{-\frac {(\tilde{y}_n^j)^2} {2}} e^{  \sqrt{2}\tilde\alpha_n^j \tilde{y}_n^j}
\end{equation}

$j=1,2,...,25$. We have then:
\begin{equation}
\label{eq8.3}    
|\alpha_n^j>=\left(\frac {1} {\pi}\right)^{\frac {1} {4}}
e^{-\frac {(\alpha_n^j)^2} {2}} e^{-\frac {|\alpha_n^j|^2} {2}}
\int e^{-\frac {(y_n^j)^2} {2}} e^{  \sqrt{2}\alpha_n^j y_n^j}|y_n^j>dy_n^j
\end{equation}
\begin{equation}
\label{eq8.4}    
|\tilde\alpha_n^j>=\left(\frac {1} {\pi}\right)^{\frac {1} {4}}
e^{-\frac {(\tilde\alpha_n^j)^2} {2}} e^{-\frac {|\tilde\alpha_n^j|^2} {2}}
\int e^{-\frac {(\tilde{y}_n^j)^2} {2}} e^{  \sqrt{2}\tilde\alpha_n^j \tilde{y}_n^j}|\tilde{y}_n^j>d\tilde{y}_n^j
\end{equation}
A coherent state of the closed bosonic string is:
    
\[|\alpha_n^j,\tilde\alpha_n^j,p^j>=|\alpha_n^j>\otimes|\tilde\alpha_n^j>\otimes|p^j>\]
\[\biggl[\left(\frac {1} {\pi}\right)^{\frac {1} {2}}
e^{-\frac {(\alpha_n^j)^2} {2}} e^{-\frac {|\alpha_n^j|^2} {2}}
e^{-\frac {(\tilde\alpha_n^j)^2} {2}} e^{-\frac {|\tilde\alpha_n^j|^2} {2}}\]
\begin{equation}
\label{eq8.5}
\int e^{-\frac {(y_n^j)^2+
(\tilde{y}_n^j)^2} {2}} e^{  \sqrt{2}\alpha_n^j y_n^j+
  \sqrt{2}\tilde{\alpha}_n^j \tilde{y}_n^j}
|y_n^j>\otimes|\tilde{y}_n^j>dy_n^jd\tilde{y}_n^j\biggr]
\otimes|p^j>
\end{equation}
And then:
\[\psi_{\alpha\tilde\alpha_n^j}(y_n^j\tilde{y}_n^j,p^j)= \psi_{\alpha_n^j}(y_n^j)\tilde\psi_{\tilde\alpha_n^j}(\tilde{y}_n^j)\phi(p^j)=\]
\[\left[\left(\frac {1} {\pi}\right)^{\frac {1} {4}}
e^{-\frac {(\alpha_n^j)^2} {2}} e^{-\frac {|\alpha_n^j|^2} {2}}
e^{-\frac {(y_n^j)^2} {2}} e^{  \sqrt{2}\alpha_n^j y_n^j}\right]\]
\begin{equation}
\label{eq8.6} 
\left[\left(\frac {1} {\pi}\right)^{\frac {1} {4}}
e^{-\frac {(\tilde\alpha_n^j)^2} {2}} e^{-\frac {|\tilde\alpha_n^j|^2} {2}}
e^{-\frac {(\tilde{y}_n^j)^2} {2}} e^{  \sqrt{2}\tilde\alpha_n^j \tilde{y}_n^j}\right]\phi(p^j).
\end{equation}

\nd The annihilation operators for the
one-dimensional harmonic oscillator is given by 
\begin{equation}
\label{eq8.7}
\hat{a}_n^j=\frac {\hat{y}_n^j+i\hat{p}^j} {\sqrt{2}}
\end{equation}
\begin{equation}
\label{eq8.8}
\hat{\tilde{a}}_n^j=\frac {\hat{\tilde{y}}_n^j+i\hat{\tilde{p}}^j} {\sqrt{2}}
\end{equation}

In the x-representation of Quantum Mechanics, this operators are
expressed via
\begin{equation}
\label{eq8.9}
\hat{a_n}^j(y_n^j)=\frac {1} {\sqrt{2}}
\left(y_n^j+\frac {\partial} {\partial y_n^j}\right)
\end{equation}
\begin{equation}
\label{eq8.10}
\hat{\tilde{a}}_n^j(\tilde{y}_n^j)=\frac {1} {\sqrt{2}}
\left(\tilde{y}_n^j+\frac {\partial} {\partial \tilde{y}_n^j}\right)
\end{equation}
Thus,  a coherent state is defined as the eigenfunction
\begin{equation}
\label{eq8.11} 
\hat{a_n}^j(y_n^j)\psi_{\alpha_n^j}(y_n^j)=\frac {1} {\sqrt{2}}
\left(y_n^j\psi_{\alpha_n^j}(y_n^j)+\frac {\partial \psi_{\alpha_n^j}(y_n^j)} {\partial y_n^j}\right)=
\alpha_n^j \psi_{\alpha_n^j}(y_n^j)
\end{equation}
\begin{equation}
\label{eq8.12} 
\hat{\tilde{a}}_n^j(\tilde{y}_n^j)\tilde\psi_{\tilde\alpha_n^j}(\tilde{y}_n^j)=\frac {1} {\sqrt{2}}
\left(\tilde{y}_n^j\tilde\psi_{\tilde\alpha_n^j}(\tilde{y}_n^j)+\frac {\partial \tilde\psi_{\tilde\alpha_n^j}(\tilde{y}_n^j)} {\partial \tilde{y}_n^j}\right)=
\tilde\alpha_n^j \tilde\psi_{\tilde\alpha_n^j}(\tilde{y}_n^j)
\end{equation}
or, equivalently,
\begin{equation}
\label{eq8.13}
 \frac {\partial \psi_{\alpha_n^j}(y_n^j)}
{\partial y_n^j}=\left(\sqrt{2}\alpha_n^j-y_n^j\right)\psi_{\alpha_n^j}(y_n^j).
\end{equation}

\begin{equation}
\label{eq8.14}
\frac {\partial \tilde\psi_{\tilde\alpha_n^j}(\tilde{y}_n^j)} {\partial \tilde{y}_n^j}=\left(\sqrt{2}\tilde\alpha_n^j-j\tilde{y}_n^j\right)\tilde\psi_{\tilde\alpha_n^j}(\tilde{y}_n^j).
\end{equation}
We have now for Virasoro operators:
\begin{equation}
\label{eq8.15}
L_0=\frac {1} {2}\hat{\alpha}_0^2+\sum\limits_{s=1}^{\infty}
\hat{\alpha}_{-s}\cdot\hat{\alpha}_s
\end{equation}
\begin{equation}
\label{eq8.16}
\tilde{L}_0=\frac {1} {2}\hat{\tilde\alpha}_0^2+\sum\limits_{s=1}^{\infty}
\hat{\tilde\alpha}_{-s}\cdot\hat{\tilde\alpha}_s
\end{equation}

\begin{equation}
\label{eq8.17}
L_k=\frac {1} {2}\sum\limits_{s=-\infty}^{\infty}
\hat{\alpha}_{k-s}\cdot\hat{\alpha}_s\;\;\;\;;\;\;\;\;k>0
\end{equation}
\begin{equation}
\label{eq8.18}
\tilde{L}_k=\frac {1} {2}\sum\limits_{s=-\infty}^{\infty}
\hat{\tilde\alpha}_{k-s}\cdot\hat{\tilde\alpha}_s\;\;\;\;;\;\;\;\;k>0
\end{equation}
And
\begin{equation}
\label{eq8.19}
\hat{\tilde{a}}_n^j=\frac {1} {\sqrt{n}}\hat{\tilde\alpha}_n^j\;\;\;n>0
\end{equation}
\begin{equation}
\label{eq8.20}
\hat{\tilde{a}}^{+j}_n=\frac {1} {\sqrt{n}}\hat{\tilde\alpha}_{-n}^j\;\;\;n>0
\end{equation}
As for the open string we have (in the representations $|y>,|\tilde{y}>$):
\begin{equation}
\label{eq8.21}
L_0=\frac {\alpha_0^2} {2}+\sum\limits_{j=1}^3
\sum\limits_{s=1}^{\infty}s\frac {\left(y_s^j-\frac {\partial} {\partial y_{s}^j}\right)}
{\sqrt{2 }}
\frac {\left(y_s^j+\frac {\partial} {\partial y_s^j}\right)}
{\sqrt{2}}+\alpha_{-s}^0\alpha_{s0}
\end{equation}
\begin{equation}
\label{eq8.22}
\tilde{L}_0=\frac {\tilde\alpha_0^2} {2}+\sum\limits_{j=1}^3
\sum\limits_{s=1}^{\infty}s\frac {\left(\tilde{y}_s^j-\frac {\partial} {\partial \tilde{y}_{s}^j}\right)}
{\sqrt{2}}
\frac {\left(\tilde{y}_s^j+\frac {\partial} {\partial \tilde{y}_s^j}\right)}
{\sqrt{2}}+\tilde\alpha_{-s}^0\tilde\alpha_{s0}
\end{equation}
And:
\begin{equation}
\label{eq8.23}
\hat{a}_n^0=\frac {\hat{y}_n^0-i\hat{p}^0} {\sqrt{2}}
\end{equation}
\begin{equation}
\label{eq8.24}
\hat{\tilde{a}}_n^0=\frac {\hat{\tilde{y}}_n^0-i\hat{\tilde{p}}^0} {\sqrt{2}}
\end{equation}
The states coherent states turn out to be now:
\begin{equation}
\label{eq8.25} 
\psi_{\alpha_n^0}(y_n^0)=\left(\frac {1} {\pi}\right)^{\frac {1} {4}}
e^{\frac {(\alpha_n^0)^2} {2}} e^{\frac {|\alpha_n^0|^2} {2}}
e^{\frac {(y_n^0)^2} {2}} e^{- \sqrt{2}\alpha_n^0 y_n^0}.
\end{equation}
\begin{equation}
\label{eq8.26} 
\tilde\psi_{\tilde\alpha_n^0}(\tilde{y}_n^0)=\left(\frac {1} {\pi}\right)^{\frac {1} {4}}
e^{\frac {(\tilde\alpha_n^0)^2} {2}} e^{\frac {|\tilde\alpha_n^0|^2} {2}}
e^{\frac {(\tilde{y}_n^0)^2} {2}} e^{- \sqrt{2}\tilde\alpha_n^0 \tilde{y}_n^0}.
\end{equation}
And:
\begin{equation}
\label{eq8.27}    
|\alpha_n^0>=\left(\frac {} {\pi}\right)^{\frac {1} {4}}
e^{\frac {(\alpha_n^0)^2} {2}} e^{\frac {|\alpha_n^0|^2} {2}}
\int e^{\frac {(y_n^0)^2} {2}} e^{-\sqrt{2}\alpha_n^0 y_n^0}|y_n^0>dy_n^0
\end{equation}
\begin{equation}
\label{eq8.28}    
|\tilde\alpha_n^0>=\left(\frac {1} {\pi}\right)^{\frac {1} {4}}
e^{\frac {(\tilde\alpha_n^0)^2} {2}} e^{\frac {|\tilde\alpha_n^0|^2} {2}}
\int e^{\frac {(\tilde{y}_n^0)^2} {2}} e^{-\sqrt{2}\tilde\alpha_n^0 \tilde{y}_n^0}|\tilde{y}_n^0>d\tilde{y}_n^0
\end{equation}
Using (\ref{eq4.18}) we have again:

\begin{equation}
\label{eq8.29}    
|||\alpha_n^0>||=0
\end{equation}
\begin{equation}
\label{eq8.30}    
|||\tilde\alpha_n^0>||=0
\end{equation}

The complete state:

\begin{equation}
\label{eq8.31}
|\alpha,\tilde\alpha,p>=\prod\limits_{n\geq 1}
\prod\limits_{j=1}^{25}\otimes|\alpha_n^j>\otimes|\alpha_n^0>\otimes|\tilde{\alpha}_n^j>\otimes
|\tilde{\alpha}_n^0>\otimes|p>
\end{equation}
Again the coherent state has a null norm, that is
\begin{equation}
\label{eq8.32}
<\alpha,\tilde{\alpha},p^{'}|\alpha,\tilde{\alpha},p>=0
\end{equation}
And then
\begin{equation}
\label{eq8.33}
<\alpha,\tilde{\alpha},p^{'}|(L_0-I)|\alpha,\tilde{\alpha},p>=0
\end{equation}
\begin{equation}
\label{eq8.34}
<\alpha,\tilde{\alpha},p^{'}|(\tilde{L}_0-I)|\alpha,\tilde{\alpha},p>=0
\end{equation}
and
\begin{equation}
\label{eq8.35}
<\alpha,\tilde{\alpha},p^{'}|L_k|\alpha,\tilde{\alpha},p>=0
\end{equation}
\begin{equation}
\label{eq8.36}
<\alpha,\tilde{\alpha},p^{'}|\tilde{L}_k|\alpha,\tilde{\alpha},p>=0
\end{equation}
 Subtracting  (\ref{eq8.34}) to (\ref{eq8.33}) we see that the level-matching condition is satisfied:
\begin{equation}
\label{eq8.36a}
<\alpha,\tilde{\alpha},p^{'}|L_0-\tilde{L}_0)|\alpha,\tilde{\alpha},p>=0
\end{equation}
The mas of the coherent state is given by
\begin{equation}
\label{eq8.37}
M^2\delta(p-p^{'})=
<\alpha,\tilde{\alpha},p^{'}|2[2(L_0+\tilde{L}_0)-\alpha_0^2-\tilde{\alpha}_0^2-2]|\alpha,\tilde{\alpha},p>=0
\end{equation}
As a consequence
\begin{equation}
\label{eq8.38}
M^2=0
\end{equation}
The same thing happens for the closed string as for the open string. In that case we have:
\begin{equation}
\label{eq8.39}
I=\int |\alpha_n^j,\tilde{\alpha}_n^j>\frac {d\alpha_n^j} {\pi}
\frac {d\tilde{\alpha}_n^j} {\pi}<\alpha_n^j,\tilde{\alpha}_n^j|
\end{equation}
and 
\begin{equation}
\label{eq8.40}
0=\int |\alpha_n^0,\tilde{\alpha}_n^0>\frac {d\alpha_n^0} {\pi}
\frac {d\tilde{\alpha}_n^0} {\pi}<\alpha_n^0,\tilde{\alpha}_n^0|
\end{equation}
We can then redefine the coherent state as follows:
\begin{equation}
\label{eq8.41}
|\alpha,\tilde\alpha,p>=\prod\limits_{n\neq 0}
\prod\limits_{j=1}^{25}\otimes|\alpha_n^j>\otimes|\tilde{\alpha}_n^j>\otimes|p>
\end{equation}
The treatment of Virasoro constraints is similar. We give here the mass of the redefined coherent state:
\begin{equation}
\label{eq8.42}
M^2\delta(p-p^{'})=<\alpha_n\tilde{\alpha}_n,p^{'}|2[2(L_0+\tilde{L}_0)-\alpha_0^2-\tilde{\alpha}_0^2-4]
|\alpha_n,\tilde{\alpha}_n,p>
\end{equation}
Then we have:
\begin{equation}
\label{eq8.43}
M^2=4\left[\sum_{n=1}^\infty n(|\alpha_n|^2+|\tilde{\alpha}_n|^2)-2\right]
\end{equation}
In this case the level-matching condition translates into:
\begin{equation}
\label{eq8.44}
\sum_{n=1}^\infty n(|\alpha_n|^2-|\tilde{\alpha}_n|^2)=0
\end{equation}

\section{Mandel Parameter for closed bosonic string cat states}

\setcounter{equation}{0}

The Mandel parameter of closed string coherent states is calculated in the same way as in open bosonic string coherent states and the results for both right handed and left handed coherent states of the closed strings are same
\[Q_1=\tilde{Q}_1=0\]
\begin{equation}
\label{eq9.1}
Q_2 =\tilde{Q}_2=-2
\end{equation}

Following the procedure for open strings the Mandel parameter for both the sides of closed string have same result and is given as:
\begin{equation}
\label{eq9.2}
Q=\tilde{Q}=-2
\end{equation}
In this section we will calculate the cat states and the corresponding Mandel parameter for closed strings.
The even cat states of closed strings are  defined as:
\begin{equation}
\label{eq9.3}
|Cat_e(\alpha_n^j)>=C(|\alpha_n^j,\tilde\alpha_n^j>+|-\alpha_n^j,-\tilde\alpha_n^j>)
\end{equation}
As:
\begin{equation}
\label{eq9.4}
<\alpha_n^{*j},\tilde\alpha_n^{*j}|-\alpha_n^j,-\tilde\alpha_n^{j}>=e^{-2|\alpha_N^J|^2}e^{-2|\tilde\alpha_N^J|^2}
\end{equation}
We have:
\begin{equation}
\label{eq9.5}
C=\frac {e^{|\alpha_N^J|^2}e^{|\tilde\alpha_N^J|^2}} {\sqrt{2(1+e^{2|\alpha_N^J|^2}e^{2|\tilde\alpha_N^J|^2})}}
\end{equation}
In a similar way that for the open string we have:
\begin{equation}
\label{eq9.6}
a_n^j|Cat_e(\alpha_n^j,\tilde\alpha_n^j)>=\alpha_n^j\frac {e^{|\alpha_N^J|^2}e^{|\tilde\alpha_N^J|^2}} {\sqrt{2(1+e^{2|\alpha_N^J|^2}e^{2|\tilde\alpha_N^J|^2})}}\left(|\alpha_n^j,\tilde\alpha_n^j>-|-\alpha_n^j,-\tilde\alpha_n^j>\right)
\end{equation}
\begin{equation}
\label{eq9.7}
\tilde{a}_n^j|Cat_e(\alpha_n^j,\tilde\alpha_n^j)>=\tilde\alpha_n^j\frac {e^{|\alpha_N^J|^2}e^{|\tilde\alpha_N^J|^2}} {\sqrt{2(1+e^{2|\alpha_N^J|^2}e^{2|\tilde\alpha_N^J|^2})}}\left(|\alpha_n^j,\tilde\alpha_n^j>-|-\alpha_n^j,-\tilde\alpha_n^j>\right)
\end{equation}
The mean values are now
\begin{equation}
\label{eq9.8}
<a_n^{+j}a_n^j>=|\alpha_n^j|^2\frac {e^{2|\alpha_N^J|^2}e^{2|\tilde\alpha_N^J|^2}-1} {e^{2|\alpha_N^J|^2}e^{2|\tilde\alpha_N^J|^2}+1}
\end{equation}
\begin{equation}
\label{eq9.9}
<\tilde{a}_n^{+j}\tilde{a}_n^j>=|\tilde\alpha_n^j|^2\frac {e^{2|\alpha_N^J|^2}e^{2|\tilde\alpha_N^J|^2}-1} {e^{2|\alpha_N^J|^2}e^{2|\tilde\alpha_N^J|^2}+1}
\end{equation}
And:
\begin{equation}
\label{eq9.10}
<a_n^{+j}a_n^ja_n^{+j}a_n^j>=|\alpha_n^j|^2\frac {e^{2|\alpha_N^J|^2}e^{2|\tilde\alpha_N^J|^2}-1} {e^{2|\alpha_N^J|^2}e^{2|\tilde\alpha_N^J|^2}+1}+|\alpha_n^j|^4
\end{equation}

\begin{equation}
\label{eq9.11}
<\tilde{a}_n^{+j}\tilde{a}_n^j\tilde{a}_n^{+j}\tilde{a}_n^j>=|\tilde\alpha_n^j|^2\frac {e^{2|\alpha_N^J|^2}e^{2|\tilde\alpha_N^J|^2}-1} {e^{2|\alpha_N^J|^2}e^{2|\tilde\alpha_N^J|^2}+1}+|\tilde\alpha_n^j|^4
\end{equation}
We are going to calculate the normalization constant for these states.
\begin{equation}
\label{eq9.12}
|Cat_o(\alpha_n^j,\tilde\alpha_n^j)>=C(|\alpha_n^j,\tilde\alpha_n^j,\tilde\alpha_n^j>-|-\alpha_n^j,-\tilde\alpha_n^j>)
\end{equation}
As:
\begin{equation}
\label{eq9.13}
<\alpha_n^{*j},\tilde\alpha_n^{*j}|-\alpha_n^j,-\tilde\alpha_n^{j}>=e^{-2|\alpha_N^J|^2}e^{-2|\tilde\alpha_N^J|^2}
\end{equation}
We obtain:
\begin{equation}
\label{eq9.14}
C=\frac {e^{|\alpha_N^J|^2}e^{|\tilde\alpha_N^J|^2}} {\sqrt{2(e^{2|\alpha_N^J|^2}e^{2|\tilde\alpha_N^J|^2}-1)}}
\end{equation}
For the annihilation operators we have:
\begin{equation}
\label{eq9.15}
a_n^j|Cat_o(\alpha_n^j,\tilde\alpha_n^j)>=\alpha_n^j\frac {e^{|\alpha_N^J|^2}e^{|\tilde\alpha_N^J|^2}} {\sqrt{2(e^{2|\alpha_N^J|^2}e^{2|\tilde\alpha_N^J|^2}-1)}}\left(|\alpha_n^j,\tilde\alpha_n^j>+|-\alpha_n^j,-\tilde\alpha_n^j>\right)
\end{equation}
\begin{equation}
\label{eq9.16}
\tilde{a}_n^j|Cat_e(\alpha_n^j,\tilde\alpha_n^j)>=\tilde\alpha_n^j\frac {e^{|\alpha_N^J|^2}e^{|\tilde\alpha_N^J|^2}} {\sqrt{2(e^{2|\alpha_N^J|^2}e^{2|\tilde\alpha_N^J|^2}-1)}}\left(|\alpha_n^j,\tilde\alpha_n^j>+|-\alpha_n^j,-\tilde\alpha_n^j>\right)
\end{equation}
We have as a consequence:
\begin{equation}
\label{eq9.17}
<a_n^{+j}a_n^j>=|\alpha_n^j|^2\frac {e^{2|\alpha_N^J|^2}e^{2|\tilde\alpha_N^J|^2}+1} {e^{2|\alpha_N^J|^2}e^{2|\tilde\alpha_N^J|^2}-1}
\end{equation}
\begin{equation}
\label{eq9.18}
<\tilde{a}_n^{+j}\tilde{a}_n^j>=|\tilde\alpha_n^j|^2\frac {e^{2|\alpha_N^J|^2}e^{2|\tilde\alpha_N^J|^2}+1} {e^{2|\alpha_N^J|^2}e^{2|\tilde\alpha_N^J|^2}-1}
\end{equation}
And:
\begin{equation}
\label{eq9.19}
<a_n^{+j}a_n^ja_n^{+j}a_n^j>=|\alpha_n^j|^2\frac {e^{2|\alpha_N^J|^2}e^{2|\tilde\alpha_N^J|^2}+1} {e^{2|\alpha_N^J|^2}e^{2|\tilde\alpha_N^J|^2}-1}+|\alpha_n^j|^4
\end{equation}

\begin{equation}
\label{eq9.20}
<\tilde{a}_n^{+j}\tilde{a}_n^j\tilde{a}_n^{+j}\tilde{a}_n^j>=|\tilde\alpha_n^j|^2\frac {e^{2|\alpha_N^J|^2}e^{2|\tilde\alpha_N^J|^2}+1} {e^{2|\alpha_N^J|^2}e^{2|\tilde\alpha_N^J|^2}-1}+|\tilde\alpha_n^j|^4
\end{equation}
The Mandel parameter for even cat states is
\begin{equation}
\label{eq9.21}
Q_1=\frac {2|\alpha_n^j|^2} {\sinh(2(|\alpha_n^j|^2)+|\tilde\alpha_n^j|^2))}
\end{equation}
\begin{equation}
\label{eq9.22}
\tilde{Q}_1=\frac {2|\tilde\alpha_n^j|^2} {\sinh(2(|\alpha_n^j|^2)+|\tilde\alpha_n^j|^2))}
\end{equation}
For the odd cat states for both right and left handed states we have
\begin{equation}
\label{eq9.23}
Q_1=-\frac {2|\alpha_n^j|^2} {\sinh(2(|\alpha_n^j|^2+|\tilde\alpha_n^j|^2))}
\end{equation}
\begin{equation}
\label{eq9.24}
\tilde{Q}_1=-\frac {2|\tilde\alpha_n^j|^2} {\sinh(2(|\alpha_n^j|^2+|\tilde\alpha_n^j|^2))}
\end{equation}

\setcounter{equation}{0}

\section{Cat states for $\mu=0$}

For $\mu=0$, the even cat states are 
\begin{equation}
\label{eq10.1}
|Cat_e(\alpha_n^0)>=C(|\alpha_n^0,\tilde\alpha_n^0>+|-\alpha_n^0,-\tilde\alpha_n^0>)
\end{equation}
The scalar products are null again.
\begin{equation}
\label{eq10.2}
<\alpha_n^{*0},\tilde\alpha_n^{*0}|-\alpha_n^0,-\tilde\alpha_n^{0}>=0
\end{equation}
\begin{equation}
\label{eq10.3}
<\alpha_n^{*0},\tilde\alpha_n^{*0}|\alpha_n^0,\tilde\alpha_n^{0}>=0
\end{equation}
\begin{equation}
\label{eq10.4}
<-\alpha_n^{*0},-\tilde\alpha_n^{*0}|-\alpha_n^0,-\tilde\alpha_n^{0}>=0
\end{equation}
These integrals vanish for the following reason
\[<\alpha_n^{0*}|\alpha_n^0>=\]
\[\left(\frac {1} {\pi}\right)^{\frac {1} {2}}
e^{\frac {(\alpha_n^0)^2} {2}} e^{\frac {|\alpha_n^0|^2} {2}}
e^{\frac {(\alpha_n^{*0})^2} {2}} e^{\frac {|\alpha_n^0|^2} {2}}\]
\begin{equation}
\label{eq10.5}
\int e^{\frac {(y_n^0)^2+(y_n^{'0})^2} {2}} 
e^{-\sqrt{2}[\alpha_n^0 y_n^0+\alpha_n^{*0} y_n^{'0}]}<y_n^{'0}||y_n^0>dy_n^0dy_n^{'0}
\end{equation}
We use:
\[<y_n^0|y_n^{'0}>=\delta(y_n^0-y_n^{'0})\]
And then:
\[<\alpha_n^{0*}|\alpha_n^0>=\]
\[\left(\frac {1} {\pi}\right)^{\frac {1} {2}}
e^{\frac {(\alpha_n^0)^2} {2}} 
e^{\frac {(\alpha_n^{*0})^2} {2}} e^{|\alpha_n^0|^2}\]
\begin{equation}
\label{eq10.6}
\int e^{(y_n^0)^2} 
e^{-\sqrt{2}[[\alpha_n^0+\alpha_n^{*0}] y_n^{0}]}dy_n^0
\end{equation}
Completing  squares
\[(y_n^0)^2-\sqrt{2}(\alpha_n^{0*}+\alpha_n^0)=
\left((y_n^0)-\frac {(\alpha_n^{0*}+\alpha_n^0)} {\sqrt{2}}\right)^2-
\frac {(\alpha_n^{0*}+\alpha_n^0)^2} {2}\]
and making the variable change
\[z=(y_n^0)-\frac {(\alpha_n^{0*}+\alpha_n^0)} {\sqrt{2}}\]
We have:
\begin{equation}
\label{eq10.7}
<\alpha_n^{0*}|\alpha_n^0>=
2\pi^{-1/2}\int\limits_0^\infty e^{z^2}dz
\end{equation}
We made the change of variable:$z^2=u$
\begin{equation}
\label{eq10.8}
<\alpha_n^{0*}|\alpha_n^0>=\pi^{-1/2}\int\limits_0^\infty u^{-1/2} e^{u}du
\end{equation}
And after: $s=1/u$
\begin{equation}
\label{eq10.9}
<\alpha_n^{0*}|\alpha_n^0>=
\pi^{-1/2}\int\limits_0^\infty s^{-3/2} e^{\frac {1} {s}}ds
\end{equation}
Using again (\ref{eq4.18}) we obtain:

\begin{equation}
\label{ep10.10}
<\alpha_n^{0*}|\alpha_n^0>=0
\end{equation}
Since in the integrals from \ref{eq8.26} to  (\ref{eq8.28}) we have the product of the type of (\ref{eq8.29}). thus all these vanish. keeping in view the values of the above expressions we evaluate
$Q$ parameter and obtain
\begin{equation}
\label{eq10.11}
Q_2=\tilde{Q}_2=-2
\end{equation}
By similar procedure the Mandel parameter is $-2$ for odd cat states.
The Mandel parameter for even cat states is
\begin{equation}
\label{eq10.12}
Q=\tilde{Q}=\frac {2|\alpha_n^j|^2} {\sinh(2(|\alpha_n^j|^2)+|\tilde\alpha_n^j|^2))}-2
\end{equation}
and for odd:
\begin{equation}
\label{eq10.13}
Q=\tilde{Q}=-\frac {2|\alpha_n^j|^2} {\sinh(2(|\alpha_n^j|^2)+|\tilde\alpha_n^j|^2))}-2
\end{equation}

\section{Coherent States in the light cone quantization formalism}

\setcounter{equation}{0}
Heuristic at par with the covariant formalism but pedagogically more intuitive, the method of light cone quantization, has been successful in advanced problems like the calculation of string amplitudes. The idea behind the light cone quantization lies actually in the fact that it preserves the physical degrees of freedom and eliminate the part of the string degrees of freedom by using residual symmetry and fix the residual gauge by setting 
$X^{+}=x^{+}+p^{+}\tau$ and thus Lorentz covariance is explicitly broken. Here the zero modes $x^+$ and $p^+$ are the integration constants, are arbitrary and left as free parameters. This choice of gauge is advantageous because it allows every point on the string to be at the same value of time or we can say that $X^+$ is independent of $\sigma$. Classically interpreting this gauge fixing means to set the oscillator coefficients $\alpha_n^+$ to zero for $n\ne0$.
The mode expansion scheme of the oscillator remains the same, but the formalism involves in setting an infinite set of modes to zero thus everything is formulated in terms of the transverse oscillators alone. Thus $\tilde\alpha_n^-$ is expressed as

\begin{equation}
\label{11.1}
\tilde\alpha_n^-=\frac{1}{2lp^{+}}\sum\limits_{k}\overline\alpha_{n-k}.\overline\alpha_{k}
\end{equation}
The quantization takes place in the same way as in canonical formalism but for the transverse oscillators only. \\
$\tilde\alpha_n^-$ are defined as operators as

\begin{equation}
\label{11.2}
\tilde\alpha_n^-=\frac{1}{2lp^{+}}\sum\limits_{k}:\overline\alpha_{n-k}.\overline\alpha_{k}:-a\delta_{n,0}
\end{equation}
For the zero mode $n=0$ the square mass operator is defined as

\begin{equation}
\label{11.3}
M^2=\frac{1}{\alpha^{'}}\left(\sum\limits_{n=1}^\infty
\overline\alpha_{-n}.\overline\alpha_{n}-a\right)
\end{equation}
This is the same mass-shell condition as found in that of the covariant treatment, with the only difference that in light cone treatment only the transverse oscillators contribute. For the theory to be Lorentz invariant, the parameter $a$ must be equal to $1$ and the dimension $D$ has to be $26$. As per the status of physical states is concerned, the light cone formalism is considered to be ghost free.\cite{gsw}
Thus the treatment of coherent states in the light cone quantization is similar to that of coherent states of open bosonic string with the only difference that $j=2,3,4,...25$. Following the same mathematical procedure we express the coherent states for the open string in light cone treatment as
\begin{equation}
\label{eq11.4}
|\alpha,p>=\prod\limits_{n\neq 0}\prod\limits_{j=2}^{25}\otimes|\alpha_n^j>\otimes|p>
\end{equation}
and for the closed string we again have the result expressed in two degrees of freedom:
\begin{equation}
\label{eq11.5}
|\alpha,\tilde\alpha,p>=\prod\limits_{n\neq 0}
\prod\limits_{j=2}^{25}\otimes|\alpha_n^j>\otimes|\tilde{\alpha}_n^j>\otimes|p>
\end{equation} 
Thus we have constructed the coherent states of both open and closed string in light cone quantization formalism and found the results are similar as that obtained from the old covariant formalism. The mass of the string will also be similar as that of the results obtained in covariant treatment. Note that the only difference between
(\ref{eq4.30}) and (\ref{eq11.4}) and between (\ref{eq8.41}) and (\ref{eq11.5}) are the states with $ j = 1 $. This clearly justifies that for practical purposes the coherent states obtained with quantization in the light cone
can be used as an approximation to the coherent states obtained in covariant form

\setcounter{equation}{0}

\section{Conclusion}

\nd 
In this paper we have obtained the Glauber coherent states for the bosonic string in the four instances, two using old covariant formalism and the two in the light cone quantization scheme, followed in this paper. In the covariant quantization formalism, we followed the original definition of coherent states of harmonic oscillator and rigorously defined the coherent states for both open and closed string. The coherent states thus obtained in both the cases, identically satisfied the Virasoro constraints at mean value. The interesting point to see here is that, the coherent states have zero mass very similar to that of the Glauber's coherent states for the electromagnetic fields, which validates the covariant quantization approach followed by us in establishing the coherent states of the string. Taking advantage of the fact that the identity resolution for the temporal components of the coherent state is null, we were able to redefine them into the  coherent states of non-zero mass.\\
 The behavior of the coherent states obtained by light cone quantization treatment is similar as that of the redefined states obtained using the covariant formalism. Thus we have a good indication of the validity of the covariant approach followed in the paper.We evaluated the Mandel parameter for the coherent states and the respective cat states to see the statistical nature of their probability distributions, which turned out to be sub-Poissonian, Poissonian and super-Poissonian depending on the value of the parameter being negative, null and positive respectively.

\nd {\bf An important fact we found is as :
Looking closely for example at (\ref{eq4.36}), we can see and infer that by a suitable choice of the $\alpha_n$ (say for example $\alpha_n=0$ ), we come across with positive norm states bearing imaginary mass, i.e the states with $ <\alpha,p^{'}|\alpha,p>=\delta(p^{'}-p)$ and $M^2<0$ which conclusively  corresponds to a tachyonic state with positive norm.}

\nd The decision to construct the coherent states of strings analogous to that of the quantum harmonic oscillator was not liked by our contemporaries, but the rhythmic mathematical treatment and the musical results of the theory is really an encouraging and motivating force to proceed to construct and establish the super coherent states in very near time. 

\section{Acknowledgement}
\nd A special thanks to our friends Mir Faizal and Salman Wani for their very helpful and in-depth discussions on this work..

\section{Conflicts of Interest Statement}
The authors certify that they have NO affiliations with or involvement in any
organization or entity with any financial interest (such as honoraria; educational grants; participation in speakers' bureaus;
membership, employment, consultancies, stock ownership, or other equity interest; and expert testimony or patent-licensing
arrangements), or non-financial interest (such as personal or professional relationships, affiliations, knowledge or beliefs) in
the subject matter or materials discussed in this manuscript.

\newpage

\renewcommand{\thesection}{\Alph{section}}

\renewcommand{\theequation}{\Alph{section}.\arabic{equation}}

\appendix

\section{Appendix: Coherent states}

The well known annihilation operator is expressed as
\begin{equation}
\label{eqa1}
\hat{a}=\frac {\hat{x}+i\hat{p}} {\sqrt{2}}
\end{equation}
In the x-representation of Quantum Mechanics, this operator is
written as:
\begin{equation}
\label{eqa2}
\hat{a}(x)=\frac {1} {\sqrt{2}}
\left(x+\frac {d} {dx}\right)
\end{equation}
By definition,  a coherent state is an eigenfunction of $\hat{a}$
\begin{equation}
\label{eqa3} 
\hat{a}(x)\psi_{\alpha}(x)=\frac {1} {\sqrt{2}}
\left(x\psi_{\alpha}(x)+\frac {d\psi_{\alpha}(x)} {dx}\right)=
\alpha \psi_{\alpha}(x)
\end{equation}
The solution of (\ref{eqa3}) reads
\begin{equation}
\label{eqa5}
 \psi_{\alpha}(x)=Ce^{-\frac {x^2} {2}}
e^{\sqrt{2}\alpha x}
\end{equation}
To evaluate the constant $C$ we use the normalization.
\begin{equation}
\label{eqa6}
 \int\limits_{-\infty}^{\infty}|\psi_{\alpha}(x)|^2dx=
|C|^2\int\limits_{-\infty}^{\infty} e^{-x^2}
e^{\sqrt{2}(\alpha+{\alpha}^{\ast}) x}dx=1
\end{equation}
We have then:
\begin{equation}
\label{eqa7}
 \int\limits_{-\infty}^{\infty}|\psi_{\alpha}(x)|^2dx=
|C|^2e^{\frac {(\alpha+{\alpha}^{\ast})^2} {2}}
\int\limits_{-\infty}^{\infty} e^{-\left(y-\frac
{\alpha+{\alpha}^{\ast}} {\sqrt{2}} \right)^2}dy=1
\end{equation}
By recourse to the  Table \cite{gr} we obtain
\begin{equation}
\label{eqa8}
 \int\limits_{-\infty}^{\infty} e^{-\left(y-\frac
{\alpha+{\alpha}^{\ast}} {\sqrt{2}} \right)^2}dy=\sqrt{\pi}.
\end{equation}
As a consequence,
\begin{equation}
\label{eqa9} 
C=\left(\frac {1} {\pi}\right)^{\frac {1} {4}}e^{-\frac
{(\alpha+{\alpha}^{\ast})^2} {4}}.
\end{equation}
Thus,  we have for $\psi_{\alpha}(x)$ the expression
\begin{equation}
\label{eqa10} 
\psi_{\alpha}(x)=
\left(\frac {1} {\pi}\right)^{\frac {1} {4}} e^{-\frac
{(\alpha+{\alpha}^{\ast})^2} {4}} e^{-\frac {x^2} {2}}
e^{\sqrt{2}\alpha x}
\end{equation}
or, equivalently,
\begin{equation}
\label{eqa11}
 \psi_{\alpha}(x)=\left(\frac {1} {\pi}\right)^{\frac {1} {4}}
 e^{i\alpha_R\alpha_I}
e^{-\frac {\alpha^2} {2}} e^{-\frac {|\alpha|^2} {2}}
e^{-\frac {x^2} {2}} e^{\sqrt{2}\alpha x}
\end{equation}
where we have:$\alpha=\alpha_R+i\alpha_I$. As $e^{i\alpha_R\alpha_I}$ is
an imaginary phase, it can be eliminated from (\ref{eqa11}) to
finally obtain
\begin{equation}
\label{eqa12} 
 \psi_{\alpha}(x)=\left(\frac {1} {\pi}\right)^{\frac {1} {4}}
e^{-\frac {\alpha^2} {2}} e^{-\frac {|\alpha|^2} {2}}
e^{-\frac {x^2} {2}} e^{\sqrt{2}\alpha x}.
\end{equation}
In the Abstract Hilbert space we have
\begin{equation}
\label{eqa13}    
|\alpha>=\left(\frac {1} {\pi}\right)^{\frac {1} {4}}
e^{-\frac {\alpha^2} {2}} e^{-\frac {|\alpha|^2} {2}}
\int e^{-\frac {x^2} {2}} e^{\sqrt{2}\alpha x}|x>dx.
\end{equation}
Thus we have obtained a simple formula for coherent states without the use of a series\\

\nd We will prove now that (\ref{eqa13}) is a Glauber's coherent state.
\nd The $n-$th HO eigenfunction is
\begin{equation} 
\label{eqb1}
\phi_n(x)={\cal H}_n\left(x\right),
\end{equation}
where ${\cal H}_n$ is Hermite's $n-$th order generalized function
\begin{equation}
\label{eqb2} 
{\cal H}_n(x)=\left(\pi^{\frac {1} {2}}2^n
n!\right)^{- \frac {1} {2}} e^{-\frac {x^2} {2}} H_n(x),
\end{equation}
Here  $H_n$ is the $n-$th Hermite polynomial. 
We take  Eq. (\ref{eqa12}), expand it in a Hermite series, and verify that one arrives to the Glauber expansion for a coherent state.

\nd In the
x-representation, the coherent state  (\ref{eqa12}) expanded into HO eigenfunctions is written:
\begin{equation}
\label{eqb3} 
\psi_{\alpha}(x)= \left(\frac {1} {\pi}\right)^{\frac {1} {4}}
e^{-\frac {\alpha^2} {2}} e^{-\frac {|\alpha|^2} {2}}
e^{-\frac {x^2} {2}} e^{\sqrt{2}\alpha x} =
\sum\limits_{n=0}^{\infty}a_n\phi_n(x),
\end{equation}
Then we express $a_n$ as:
\begin{equation}
\label{eqb4} 
a_n=\int \psi_{\alpha}(x)\phi_n(x) dx,
\end{equation}
We can further write
\begin{equation}
\label{eqb5} 
a_n=\left(\frac {1} {\pi}\right)^{\frac {1} {4}}
e^{-\frac {\alpha^2} {2}} e^{-\frac {|\alpha|^2} {2}}
\int e^{-\frac {x^2} {2}} e^{\sqrt{2}\alpha x}\phi_n(x)dx.
\end{equation}
We obtain:
\begin{equation}
\label{eqb6} 
a_n=\pi^{-\frac {1} {4}}
e^{-\frac {\alpha^2} {2}}
e^{-\frac {|\alpha|^2} {2}} \int\limits_{-\infty}^{\infty}
e^{-\frac {x^2} {2}} e^{\sqrt{2}\alpha x} 
{\cal H}_n\left(x\right)dx.
\end{equation}
We can also write it as:
\begin{equation}
\label{eqb7} 
a_n=\pi^{-\frac {1} {4}}
\left(\pi^{\frac {1} {2}}2^n n!\right)^{- \frac {1} {2}} 
e^{-\frac {\alpha^2} {2}}
e^{-\frac {|\alpha|^2} {2}} \int\limits_{-\infty}^{\infty}
e^{-x^2} e^{\sqrt{2}\alpha x} 
H_n\left(x\right)dx.
\end{equation}
or equivalently in we express it as:
\begin{equation}
\label{eqb8} 
a_n= \frac {\pi^{-\frac {1} {4}} e^{-\frac
{|\alpha|^2} {2}}} {\left(n!2^n\pi^{\frac {1} {2}}\right)^{ \frac
{1} {2}}} \int\limits_{-\infty}^{\infty} e^{-\left(y-\frac
{\alpha} {\sqrt{2}}\right)^2} H_n(y)\;dy.
\end{equation}
We appeal now to a result of (see \cite{gr}) to obtain
\begin{equation}
\label{eqb9} 
a_n= \frac {\pi^{-\frac {1} {4}} e^{-\frac
{|\alpha|^2} {2}}} {\left(n!2^n\pi^{\frac {1} {2}}\right)^{ \frac
{1} {2}}} \pi^{\frac {1} {2}}2^{\frac {n} {2}}\alpha^n,
\end{equation}
and
\begin{equation}
\label{eqb10} 
a_n=\frac {\alpha^n} {\sqrt{n!}}e^{-\frac {|\alpha|^2} {2}}.
\end{equation}
Replacing (\ref{eqb10}) in (\ref{eqb3}) we obtain the Glauber's result:
\begin{equation}
\label{eqb11} 
\psi_\alpha(x)=e^{-\frac {|\alpha|^2} {2}}\sum\limits_{n=0}^\infty\frac {\alpha^n} {\sqrt{n!}}\phi_n(x).
\end{equation}
In the Hilbert's abstract space (\ref{eqb11}) reads:
\begin{equation}
\label{eqb12} 
|\alpha>=e^{-\frac {|\alpha|^2} {2}}\sum\limits_{n=0}^\infty\frac {\alpha^n} {\sqrt{n!}}|n>
\end{equation}
We have proved that (\ref{eqa13}) and (\ref{eqb12}) are equals.

\setcounter{equation}{0}

\section{Appendix: Analytic extension}

We consider the integral:
\begin{equation}
\label{eqc1}
\int\limits_0^{\infty}x^{\nu-1}(x+\gamma)^{\mu-1}
e^{-\frac {\beta} {x}}dx=\beta^{\frac {\nu-1} {2}}
\gamma^{\frac {\nu-1} {2}+\mu}
\Gamma(1-\mu-\nu)
e^{\frac {\beta} {2\gamma}}
W_{\frac {\nu-1} {2}+\mu,-\frac {\nu} {2}}\left(
\frac {\beta} {\gamma}\right),
\end{equation}
$|\arg(\gamma)|<\pi$, $\Re(1-\mu-\nu)>0$, where
 $W$ is the second  Whittaker's function. 
Note that $\Re(\beta)>0$ is not required, as made clear by Gradshteyn and Rizhik in their table 
\cite{gr} (this formula  figure in page 340, eq. (7), called ET II 234(13)a, where reference is made  to  \cite{gr1} (Caltech's 
Bateman Project).  The last letter ''a'' indicates that analytical extension has been performed. 
Selecting  $\mu=1$ in (\ref{eqc1}) we obtain:
\begin{equation}
\label{eqc2}
\int\limits_0^{\infty}x^{\nu-1}
e^{-\frac {\beta} {x}}dx=\beta^{\frac {\nu-1} {2}}
\gamma^{\frac {\nu+1} {2}}\Gamma(-\nu)e^{\frac {\beta} {2\gamma}}
W_{\frac {\nu+1} {2},-\frac {\nu} {2}}\left(\frac {\beta}{\gamma}\right),
\end{equation}
The last formula is valid for $\nu\neq 0,1,2,3,.....$
The following formula appears in the same table \cite{gr}:
\begin{equation}
\label{eqc3}
W_{\frac {\nu+1} {2},-\frac {\nu} {2}}\left(\frac {\beta}{\gamma}\right)=M_{\frac {\nu+1} {2},\frac {\nu} {2}}\left(
\frac {\beta} {\gamma}\right)=\left(\frac {\beta} {\gamma}\right)^{\frac {\nu+1} {2}}e^{-\frac {\beta} {2\gamma}},
\end{equation}
where  $M$ is the first Whittaker's function. We have then the result:
 \begin{equation}
\label{eqc4}
\int\limits_0^{\infty}x^{\nu-1}e^{-\frac {\beta} {x}}dx=\beta^{\nu}\Gamma(-\nu)
\end{equation}
This integral can be evaluated using the generalization of the Bollini and Giambiagi dimensional regularization obtained \cite {jpco} for $\nu =1,2,3,... $ Changing now $\beta$ by $-\beta$ in (\ref{eqc4}) we have
\begin{equation}
\label{eqc5}
\int\limits_0^{\infty}x^{\nu-1}e^{\frac {\beta} {x}}dx=(-\beta)^{\nu}\Gamma(-\nu)
\end{equation}
Eq.  (\ref{eqc5}) displays a cut at 
$\Re(\beta)>0$.\\
 One can therefore choose \\ 
 $(-\beta)^\nu=e^{i\pi\nu}\beta^\nu$, $(-\beta)^\nu=e^{-i\pi\nu}\beta^\nu$, or $(-\beta)^\nu=\cos(\pi\nu)\beta^\nu$.\\
As the integral must be real for $\beta$ to be real, we choose the last possibility of the three expressions available and finally obtain:
\begin{equation}
\label{eqc6}
\int\limits_0^{\infty}x^{\nu-1}
e^{\frac {\beta} {x}}dx=\cos(\pi\nu)\beta^{\nu}
\Gamma(-\nu),
\end{equation}
We have used this important result in section 4 of this paper.

\setcounter{equation}{0}

\section{Appendix: The new Glauber theory}

\nd We will show  now that (\ref{eq4.15}) is a new type Glauber's coherent state.
\nd In the
x-representation, the new coherent state  (\ref{eq4.15}) expanded into HO eigenfunctions is written:
\begin{equation}
\label{eqd1} 
\psi_{\alpha}(x)= \left(\frac {1} {\pi}\right)^{\frac {1} {4}}
e^{\frac {\alpha^2} {2}} e^{\frac {|\alpha|^2} {2}}
e^{\frac {x^2} {2}} e^{-\sqrt{2}\alpha x} =
\sum\limits_{n=0}^{\infty}a_n\phi_n(x),
\end{equation}
Then we express $a_n$ as:
\begin{equation}
\label{eqd2} 
a_n=\int \psi_{\alpha}(x)\phi_n(x) dx,
\end{equation}
We can further write
\begin{equation}
\label{eqd3} 
a_n=\left(\frac {1} {\pi}\right)^{\frac {1} {4}}
e^{\frac {\alpha^2} {2}} e^{\frac {|\alpha|^2} {2}}
\int e^{\frac {x^2} {2}} e^{-\sqrt{2}\alpha x}\phi_n(x)dx.
\end{equation}
We obtain:
\begin{equation}
\label{eqd4} 
a_n=\pi^{-\frac {1} {4}}
e^{\frac {\alpha^2} {2}}
e^{\frac {|\alpha|^2} {2}} \int\limits_{-\infty}^{\infty}
e^{\frac {x^2} {2}} e^{-\sqrt{2}\alpha x} 
{\cal H}_n\left(x\right)dx.
\end{equation}
We can also write it as:
\begin{equation}
\label{eqd5} 
a_n=\pi^{-\frac {1} {4}}
\left(\pi^{\frac {1} {2}}2^n n!\right)^{- \frac {1} {2}} 
e^{\frac {\alpha^2} {2}}
e^{\frac {|\alpha|^2} {2}} \int\limits_{-\infty}^{\infty}
e^{-\sqrt{2}\alpha x} 
H_n\left(x\right)dx.
\end{equation}
or equivalently in we express it as:
\begin{equation}
\label{eqd6} 
\int\limits_{-\infty}^{\infty}
e^{i\alpha x} 
H_n\left(x\right)dx=\tilde{H}_n(\alpha)
\end{equation}
Here $\tilde{H}_n(\alpha)$ is the complex Fourier transform of $H_n\left(x\right)$,
and it is an analytic functional \cite{gue}
\begin{equation}
\label{eqd7} 
a_n=\pi^{-\frac {1} {4}}
\left(\pi^{\frac {1} {2}}2^n n!\right)^{- \frac {1} {2}} 
e^{\frac {\alpha^2} {2}}
e^{\frac {|\alpha|^2} {2}} \tilde{H}_n(i\sqrt{2}\alpha)
\end{equation}
Replacing (\ref{eqd7}) in (\ref{eqd1}) we obtain the new Glauber's result:
\begin{equation}
\label{eqd8} 
\psi_\alpha(x)=\pi^{-\frac {1} {4}}
e^{\frac {\alpha^2} {2}}
e^{\frac {|\alpha|^2} {2}}
\sum\limits_{n=0}^\infty\left(\pi^{\frac {1} {2}}2^n n!\right)^{- \frac {1} {2}} 
\tilde{H}_n(i\sqrt{2}\alpha)\phi_n(x)
\end{equation}
In the Rigged Hilbert Abstract Space (\ref{eqd8}) reads:
\begin{equation}
\label{eqd9} 
|\alpha>=\pi^{-\frac {1} {4}}
e^{\frac {\alpha^2} {2}}
e^{\frac {|\alpha|^2} {2}}
\sum\limits_{n=0}^\infty\left(\pi^{\frac {1} {2}}2^n n!\right)^{- \frac {1} {2}} 
\tilde{H}_n(i\sqrt{2}\alpha)|n>
\end{equation}

\end{document}